\documentclass[twocolumn,pre,twoside,floatfix]{revtex4}

\usepackage{amsmath}
\usepackage{amssymb}
\usepackage{graphicx}
\usepackage{dcolumn}
\usepackage{bm}
\usepackage{color}

\def\dps{\displaystyle}

\def\Eq#1{(\ref{eq:#1})}
\def\d{\mathrm{d}}

\def\epsilon{\varepsilon}
\def\theta{\vartheta}
\def\rho{\varrho}

\def\vec#1{\mathbf{#1}}

\begin{document}


\title{Non-equilibrium interfacial tension during relaxation}

\author{Markus Bier}
\email{bier@is.mpg.de}
\affiliation
{
   Max-Planck-Institut f\"ur Intelligente Systeme, 
   Heisenbergstr.\ 3,
   70569 Stuttgart,
   Germany, 
   and
   Institut f\"ur Theoretische Physik IV,
   Universit\"at Stuttgart,
   Pfaffenwaldring 57,
   70569 Stuttgart,
   Germany
}

\date{3 October 2015}

\begin{abstract}
The concept of a non-equilibrium interfacial tension, defined via the work
required to deform the system such that the interfacial area is changed while
the volume is conserved, is investigated theoretically in the context of the
relaxation of an initial perturbation of a colloidal fluid towards the 
equilibrium state.
The corresponding general formalism is derived for systems with planar 
symmetry and applied to fluid models of colloidal suspensions and polymer 
solutions.
It is shown that the non-equilibrium interfacial tension is not necessarily
positive, that negative non-equilibrium interfacial tensions are consistent
with strictly positive equilibrium interfacial tensions, and that the sign of
the interfacial tension can influence the morphology of density perturbations
during relaxation.
\end{abstract}

\maketitle


\section{\label{sec:intro}Introduction}

The equilibrium interfacial tension between two coexisting fluid phases is a 
well established concept of enormous relevance to numerous research areas such
as capillarity, wetting, the morphology of drops and bubbles, interface 
dynamics, and the adsorption of surfactant molecules or colloidal particles at
interfaces \cite{Safran2003,deGennes2004}. 
It is well-known that the equilibrium interfacial tension can be equivalently
viewed either mechanically in terms of the difference of the normal and the 
transversal pressure tensor components or thermodynamically via the work 
required to reversibly deform the system in such a way that the interfacial 
area is changed while the system volume is preserved \cite{Rowlinson2002}.
Both interpretations require the notion of the position of the interface, 
i.e., of the non-uniform part of the system,
which can be defined via the concept of a Gibbs dividing interface
\cite{Rowlinson2002}.
However, it can be shown that the value of the equilibrium interfacial tension
between two fluid phases is independent of the choice of the Gibbs dividing
interface \cite{Rowlinson2002}.

In contrast, the general concept of an interfacial tension in non-equilibrium 
situations, e.g., a non-equilibrium interface between two coexisting
equilibrium bulk phases or an interface between two bulk phases being not
at coexistence with each other, has not been uniquely established yet.
Several attempts have been made to extend the notion of an interfacial tension
to non-equilibrium systems in order to interpret dynamic phenomena at 
interfaces \cite{Joos1999,Drelich2006}.
The approaches range from time-resolved interfacial tension measurements 
\cite{Peach1996} via integrations of the Gibbs equation \cite{Maze1971,
Adamczyk1987,Millner1994,Eastoe2000} to capillary wave analyses
\cite{Cicuta2000}.
The overall assumption underlying all these non-equilibrium interfacial 
tensions is that of a sufficiently weakly perturbed equilibrium interface.
In fact, it has been shown recently that the interfacial structure between 
two fluid bulk phases not in thermodynamic equilibrium with each other 
converges rapidly towards that of the equilibrium interface between the two
coexisting bulk phases at the same temperature \cite{Bier2013}.
Hence, different notions of dynamic interfacial tensions can be expected
to coincide quantitatively provided they do so for equilibrium interfaces.
However, unlike for equilibrium interfaces, the value of the non-equilibrium
interfacial tension depends on the choice of the Gibbs dividing interface
\cite{Levitas2014}.

It is well-known that the interfacial tension of equilibrium fluid-fluid
interfaces is strictly positive because otherwise the interface would be
unstable with respect to capillary wave fluctuations.
The aim of the present work is to demonstrate that this is not necessarily
the case for non-equilibrium interfacial tensions.
To that end, the relaxation of an initial 
non-uniformity 
of a fluid deep inside the one-phase region of the phase diagram is considered.
Under these conditions
the equilibrium state is uniform, i.e.,
no equilibrium interface exists.
Therefore, the initial non-uniformity, i.e., the interface, 
is a purely non-equilibrium structure,
which ultimately vanishes during relaxation towards equilibrium.

The present analysis is based on the notion of a non-equilibrium interfacial
tension similar to the thermodynamic definition related to the work of 
reversible isochoric deformations.
It has been shown in Ref.~\cite{Bier2013} that this notion of a non-equilibrium
interfacial tension is not only consistent with the equilibrium interfacial
tension but also useful in the context of interfaces between phases separated
by a fluid-fluid phase transition.
In order to avoid the difficulty due to the dependence of the non-equilibrium
interfacial tension on the choice of the Gibbs dividing interface, setups
are considered for which the cross-sections of the system perpendicular to
one spatial direction ($z$-axis) are all congruent.
For such setups the number density profiles of the fluid vary only along the
$z$-axis and the interfacial area equals the area of any of the congruent 
cross-sections.
Then locating the interface position is not necessary.

In order to apply the definition of the non-equilibrium interfacial tension
as the work due to a reversible isochoric deformation in practice, the 
deformation has to be performed faster than the relaxation of the fluid under 
consideration.
However, this is not possible in practice for simple fluids, since deformations
have to be induced by deformations of the container walls, from where they 
propagate into the fluid via particle-particle interactions.
Hence reversible deformations of a simple fluid cannot be achieved on time 
scales faster than the relaxations.

One way out of this dilemma is to consider fluids of colloidal particles
which are dispersed in a molecular solvent.
In that situation, deformations of the container walls induce deformations
of the molecular solvent which propagate on time scales much shorter than
the relaxation time of the colloids. 
Hence the solvent is the tool that exerts the force generated by the
container walls onto the colloidal particles, which are considered as the
fluid to be studied.

The general formalism used in this work is derived in Sec.~\ref{sec:formalism}.
First some notation to model colloidal suspensions are introduced in
Sec.~\ref{subsec:colloids}.
The non-equilibrium properties as well as the interfacial structures of
colloidal fluids can be described simultaneously within dynamic density 
functional theory (DDFT) \cite{Evans1979,Dieterich1990,Kawasaki1994,Dean1996,
MariniBettoloMarconi1999,MariniBettoloMarconi2000}, the relevant concepts 
of which are summarized in Sec.~\ref{subsec:ddft}.
The expression of the interfacial tension as a functional of the interfacial
density profile is derived in detail in Sec.~\ref{subsec:tension}.
The evaluation of the interfacial tension for temporally evolving density 
profiles obtained within DDFT leads to the time-dependence of the interfacial
tension in Sec.~\ref{subsec:timedependence}.
The the application of the general formalism of Sec.~\ref{sec:formalism} is 
illustrated in Sec.~\ref{sec:applications} for three realistic fluid models,
all of which exhibit negative values of the non-equilibrium interfacial tension.
Conclusions from the possibility of negative non-equilibrium interfacial 
tensions and the property of strictly positive equilibrium interfacial tensions
are drawn in Sec.~\ref{sec:conclusions}.


\section{\label{sec:formalism}General formalism}

\subsection{\label{subsec:colloids}Colloidal suspensions}

Consider a three-dimensional monodisperse suspension of colloidal particles
which interact amongst each other via an isotropic pair potential $U(r), 
r=|\vec{r}_1-\vec{r}_2|$.
The temperature $T$ and the mean number density $\bar{\rho}$ of colloidal
particles are chosen such that in the absence of an external field the
equilibrium state of the system is that of a uniform isotropic fluid  well
inside a one-phase region of the phase diagram.

The fluid structure of that uniform equilibrium state is given by the isotropic
direct correlation function $\bar{c}(r):=c(r,\bar{\rho}),
r=|\vec{r}_1-\vec{r}_2|$.
It can be shown very generally that if $U(r)$ decays to zero for $r\to\infty$ 
faster than some power law $\sim r^{-s}, s>5$, i.e., if there exists some $s>5$
with $\dps\lim_{r\to\infty} r^sU(r) = 0$, the three-dimensional Fourier 
transform of $\bar{c}(r)$ is given by
\begin{align}
   \widehat{\bar{c}}(q) = 
   \widehat{\bar{c}}_0 + \widehat{\bar{c}}_2q^2 + \mathcal{R}(q)
   \label{eq:cq}   
\end{align}
with $q=|\vec{q}|$ and with $\mathcal{R}(q)$ decaying to zero for $q\to0$ 
faster than $q^2$; the derivative $\mathcal{R}'(q)$ of $\mathcal{R}(q)$ with 
respect to $q$ turns out to decay to zero for $q\to0$ faster than $q$.
The situation just described covers practically all cases of isotropically
interacting colloids, e.g., non-retarded dispersion forces in dense 
suspensions, retarded dispersion forces in dilute suspensions, orientationally
averaged magnetic dipole interaction, as well as pure screened Coulomb 
interaction for suspensions with the index of refraction of the solvent being
matched to that of the colloids.

At time $t<0$ a static external field is applied to the colloidal suspension
giving rise to a non-uniform equilibrium number density profile 
$\rho_0(\vec{r})$ with mean number density $\dps\bar{\rho}=\frac{1}{|V|}\int_V
\!\d^3r\,\rho_0(\vec{r})$.
At time $t=0$ the external field is switched off leaving the colloidal 
suspension in a non-equilibrium state, which relaxes for $t\to\infty$ towards
the equilibrium state with the uniform number density $\bar{\rho}$.
In the following the temporal evolution of the non-equilibrium state of the 
colloidal suspension is investigated in terms of the time-dependent number 
density profile $\rho(\vec{r},t)$, which at time $t=0$ is given by 
$\rho(\vec{r},0)=\rho_0(\vec{r})$ and which attains the long time limit $\dps
\lim_{t\to\infty}\rho(\vec{r},t)=\bar{\rho}$.
The deviation of the number density $\rho(\vec{r},t)$ from its long time limit
$\bar{\rho}$ is denoted as $\phi(\vec{r},t):=\rho(\vec{r},t)-\bar{\rho}$, which
decays to zero in the long time limit: $\dps\lim_{t\to\infty}\phi(\vec{r},t)=
0$.


\subsection{\label{subsec:ddft}Dynamic density functional theory}

The relaxation of colloidal suspensions can be described as an overdamped 
conserved dynamics (model B), which can be formulated in terms of dynamic 
density functional theory \cite{Dieterich1990}.
If the system is prepared in an arbitrary initial state $\rho(\vec{r},t=0)$,
its state $\rho(\vec{r},t>0)$ evolves with time $t$ such that the Helmholtz
free energy $F[\rho(t)]$ reaches a minimum at $t\to\infty$.

The Helmholtz free energy density functional $F[\rho]$ can be written as
\begin{align}
   \beta F[\rho] = 
   \int\!\d^3r\,\rho(\vec{r})\left(\ln(\rho(\vec{r})\Lambda^3)-1\right)
   + \beta F^\text{ex}[\rho-\bar{\rho}]
   \label{eq:betaF}
\end{align}
with the inverse temperature $\beta=1/(k_\text{B}T)$.
The excess free energy functional $\beta F^\text{ex}[\phi]$ can be expanded
in a functional Taylor series in powers of $\phi=\rho-\bar{\rho}$, which
results in the virial expansion
\begin{align}
   \beta F^\text{ex}[\phi] = 
   &\beta F^\text{ex}[0] - c^{(1)}(\bar{\rho})\int\!\d^3r\,\phi(\vec{r})
    \notag\\
   &- \frac{1}{2}\int\!\d^3r\!\int\!\d^3r'\,
    c^{(2)}(|\vec{r}-\vec{r'}|,\bar{\rho})\phi(\vec{r})\phi(\vec{r'})
    \notag\\
   &+ \mathcal{O}(\phi^3). 
    \label{eq:betaFexfull}
\end{align}
The term linear in $\phi$ drops out because $\dps\int\d^3r\,\phi(\vec{r})=0$,
and $c^{(2)}(r,\bar{\rho})=c(r,\bar{\rho})=\bar{c}(r)$.
Since $\phi(\vec{r},t)\to0$ for $t\to\infty$, the approximation
\begin{align}
   \beta F^\text{ex}[\phi] \simeq\
   &\beta F^\text{ex}[0] 
    \notag\\	
   &- \frac{1}{2}\int\!\d^3r\!\int\!\d^3r'\,\bar{c}(|\vec{r}-\vec{r'}|)
    \phi(\vec{r})\phi(\vec{r'}) 
    \label{eq:betaFex}
\end{align}
is applicable, at least at sufficiently late times $t$.

Since the colloidal processes considered in the present work, which are
described in terms of $\rho(\vec{r},t)$, are much slower than the molecular
degrees of freedom, one can assume local thermodynamic equilibrium and define
the local chemical potential \cite{Dieterich1990}
\begin{align}
   \beta\mu(\vec{r},[\rho(t)]) 
   &= \frac{\delta \beta F}{\delta \rho(\vec{r})}[\rho(t)] 
    \notag\\
   &= \ln(\rho(\vec{r},t)\Lambda^3) + \beta\mu^\text{ex}(\vec{r},[\phi(t)])
   \label{eq:mudef}
\end{align}
with
\begin{align}
   \beta\mu^\text{ex}(\vec{r},[\phi(t)])
   &= \frac{\delta \beta F^\text{ex}}{\delta \phi(\vec{r})}[\phi(t)]
    \notag\\
   &\simeq -\int\!\d^3r'\,\bar{c}(|\vec{r}-\vec{r'}|)\phi(\vec{r'},t).
   \label{eq:muexdef}
\end{align}
The local force $-\nabla\mu(\vec{r},[\rho(t)])$ generates a flux 
\cite{Dieterich1990}
\begin{align}
   \vec{j}(\vec{r},[\rho(t)]) 
   &= \beta D\left(-\nabla\mu(\vec{r},[\rho(t)])\right)\rho(\vec{r},t)
    \notag\\
   &= -D\left(\frac{\nabla\rho(\vec{r},t)}{\rho(\vec{r},t)}+
    \nabla\beta\mu^\text{ex}(\vec{r},[\phi(t)])\right)\rho(\vec{r},t)
    \notag\\
   &= -D\big(\underbrace{\nabla\rho(\vec{r},t)}_{=\nabla\phi(\vec{r},t)} 
    + \bar{\rho}\nabla\beta\mu^\text{ex}(\vec{r},[\phi(t)])\big) 
    + \mathcal{O}(\phi^2)
    \notag\\
   &\simeq -D\left(\nabla\phi(\vec{r},t)
    + \bar{\rho}\nabla\beta\mu^\text{ex}(\vec{r},[\phi(t)])\right) 
   \label{eq:jdef}
\end{align}
with the diffusion constant $D$.
From the continuity equation of the particle number,
\begin{align}
   \frac{\partial\rho(\vec{r},t)}{\partial t} = 
    -\nabla\cdot\vec{j}(\vec{r},[\rho(t)]),
   \label{eq:contirho}
\end{align}
one obtains the conserved dynamics (model B \cite{Hohenberg1977})
equation of motion of $\phi(\vec{r},t)$:
\begin{align}
   \frac{\partial\phi(\vec{r},t)}{\partial t} \simeq
   D\left(\nabla^2\phi(\vec{r},t)
    + \bar{\rho}\nabla^2\beta\mu^\text{ex}(\vec{r},[\phi(t)])\right).
   \label{eq:contiphi}
\end{align}

For the sake of simplicity only planar density profiles $\rho(\vec{r},t)=
\rho(z,t)$, and hence $\phi(\vec{r},t)=\phi(z,t)$, are considered in the
following.
Writing 
\begin{align}
   \phi(z,t) =: 
   \frac{1}{2\pi}\int\!\d q_z\,
   \widehat{\phi}(q_z,t)\exp(iq_zz)
   \label{eq:fourierphi}
\end{align}
one obtains from Eq.~\Eq{muexdef}
\begin{align}
   \beta\mu^\text{ex}(z,[\phi(t)])
   \simeq -\frac{1}{2\pi}\int\!\d q_z\,
   \widehat{\bar{c}}(|q_z|)\widehat{\phi}(q_z,t)\exp(iq_zz)
   \label{eq:foueriermuex}
\end{align}
with $\widehat{\bar{c}}(q),q=|\vec{q}|,$ denoting the three-dimensional 
Fourier-transform of $\bar{c}(r), r=|\vec{r}|,$ (see Eq.~\Eq{cq}).
Hence, the equation of motion Eq.~\Eq{contiphi} for $\widehat{\phi}(q_z,t)$
reads
\begin{align}
   \frac{\partial\widehat{\phi}(q_z,t)}{\partial t}
   \simeq -Dq_z^2\left(1-\bar{\rho}\,\widehat{\bar{c}}(|q_z|)\right)
          \widehat{\phi}(q_z,t),
   \label{eq:contiphihat}
\end{align}
which is readily solved by
\begin{align}
   \widehat{\phi}(q_z,t) 
   &= \widehat{\phi}(q_z,0)
   \exp\left(-Dq_z^2t(1-\bar{\rho}\,\widehat{\bar{c}}(|q_z|))\right)
   \notag\\
   &= \widehat{\phi}(q_z,0)
   \exp\left(-\frac{Dq_z^2t}{S(|q_z|)}\right),
   \label{eq:contisolution}
\end{align}
where
\begin{align}
   S(q) = \frac{1}{1-\bar{\rho}\,\widehat{\bar{c}}(q)}
   \label{eq:strucfac}
\end{align}
is the colloidal structure factor of the uniform suspension.
Equation~\Eq{contisolution} expresses the relaxation of the Fourier mode
$\widehat{\phi}(q_z,t)$ towards zero on the $q_z$-dependent time scale
\begin{align}
   \tau(|q_z|):= \frac{S(|q_z|)}{Dq_z^2}.
   \label{eq:timescale}
\end{align}


\subsection{\label{subsec:tension}Interfacial tension}

Consider a system with density profiles $\rho(\vec{r},t)=\rho(z,t)$, 
i.e., $\phi(\vec{r},t)=\phi(z,t)$, 
inside a container with all cross-sections perpendicular to the $z$-axis being
congruent of area $A$.
The interfacial tension is defined via the work which is required to change the
interfacial area $A$ at constant temperature, volume, and particle number 
\cite{Rowlinson2002}.
In colloidal suspensions such a change of the interfacial area can be 
achieved by deforming
the incompressible molecular solvent, which drags the colloidal particles with
it.
Due to the separation of molecular and colloidal time scales, one is able to
perform this deformation, on the one hand, sufficiently slowly in order to stay
in the regime of low Reynolds numbers to avoid dissipation due to turbulence,
and, on the other hand, sufficiently fast such that the colloidal distribution
$\rho(z,t)$ is practically not evolving during the measurement.

\begin{figure}[!t]
   \includegraphics{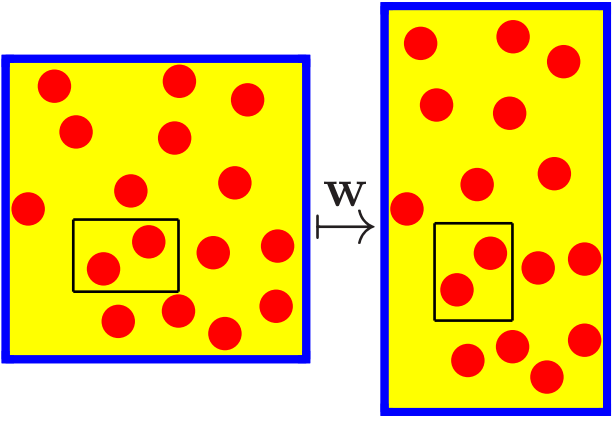}
   \caption{Sketch of the effect of the displacement field
           $\vec{w}(\vec{r})$ of the molecular solvent induced by a 
           deformation of the container walls. 
           Dispersed colloidal particles are dragged along with the
           solvent so that the volume as well as the number of contained
           colloids is conserved for each fluid element (small black frame).
           }
   \label{fig:1}
\end{figure}

By slowly deforming the boundaries of the container in a time intervall of
length $t_\text{deform}$, a flow field $\vec{v}(\vec{r},t)$ of the solvent is
induced, which fulfills the Stokes equation within the regime of low Reynolds
numbers.
Here $t_\text{deform}$ has to be long compared to the relaxation time of the
molecular solvent (typically nanoseconds) and short compared to the relaxation
time of the colloids (at least milliseconds). 
Moreover, if the velocity of the container walls is negligible compared to
the sound velocity of the solvent, no pressure gradients occur inside the
solvent so that the Stokes equation reduces to the Laplace equation 
$\nabla^2\vec{v}(\vec{r},t)=0$.
Complementary to the Eulerian description of the flow in terms of the velocity
$\vec{v}(\vec{r},t)$ at position $\vec{r}$ and time $t$ one can use the
Lagrangian description in terms of the displacement $\vec{W}(\vec{r},t)$ within
the time interval $t$ of a fluid element originally located at position 
$\vec{r}$, where both descriptions are related by
\begin{align}
   \frac{\partial \vec{W}}{\partial t}(\vec{r},t) =
   \vec{v}(\vec{r}+\vec{W}(\vec{r},t),t), \quad \vec{W}(\vec{r},0)=0.
   \label{eq:ELdescription}
\end{align}
From $\vec{W}(\vec{r},t)$ one obtains the displacement field 
$\vec{w}(\vec{r}):=\vec{W}(\vec{r},t_\text{deform})$ experienced by fluid 
elements during the time $t_\text{deform}$.
This displacement field $\vec{w}(\vec{r})$ corresponds to the deformation map
$\vec{r}\mapsto \vec{r}_\vec{w}:=\vec{r}+\vec{w}(\vec{r})$ of the 
three-dimensional space (see Fig.~\ref{fig:1}).
It is assumed that the container is much larger than the colloidal particles
so that $\vec{w}(\vec{r})$ is slowly varying on colloidal length scales.
Due to the incompressibility of the solvent, 
i.e., $\nabla\cdot\vec{v}(\vec{r},t)=0$,
\begin{align}
   \nabla\cdot\vec{w}(\vec{r})=\mathcal{O}(\|\vec{w}\|^2)
   \label{eq:divw}
\end{align}
holds for small deformations $\|\vec{w}\|\to0$.
Since the considered deformations conserve the volume as well as the number of
colloidal particles of the fluid elements (see Fig.~\ref{fig:1}), the number
densities are not changed by the deformation map: 
$\rho_\vec{w}(\vec{r}_\vec{w}) = \rho(\vec{r})$.

Upon deformation of $\phi(\vec{r})=\rho(\vec{r})-\bar{\rho}$ with the
displacement field $\vec{w}(\vec{r})$, the excess free energy in 
Eq.~\Eq{betaFex} leads to
\begin{align}
   &\beta F^\text{ex}[\phi_\vec{w}] - \beta F^\text{ex}[0]
    \notag\\	
   &\simeq -\frac{1}{2}\int\!\d^3r_\vec{w}\!\int\!\d^3r'_\vec{w}\,
    \bar{c}(|\vec{r}_\vec{w}-\vec{r'}_\vec{w}|)\phi_\vec{w}(\vec{r}_\vec{w})
    \phi_\vec{w}(\vec{r'}_\vec{w}) 
    \notag\\
   &= -\frac{1}{2}\int\!\d^3r\!\int\!\d^3r'\,
    \bar{c}(|\vec{r}-\vec{r'}+\vec{w}(\vec{r})-\vec{w}(\vec{r'})|)
    \phi(\vec{r})\phi(\vec{r'}).
    \label{eq:betaFexdeformed}
\end{align}
Writing $c^{(2)}(\vec{r}):=\bar{c}(|\vec{r}|)$, the direct correlation function 
in Eq.~\Eq{betaFexdeformed} can be expanded for small deformations:
\begin{multline}
   c^{(2)}(\vec{r}-\vec{r'}+\vec{w}(\vec{r})-\vec{w}(\vec{r'}))
    = c^{(2)}(\vec{r}-\vec{r'})
   \\
   + (w_k(\vec{r})-w_k(\vec{r'}))
   \frac{\partial c^{(2)}}{\partial r_k}(\vec{r}-\vec{r'}) +
   \mathcal{O}(\|\vec{w}\|^2),
   \label{eq:cdeformed}
\end{multline}
where an implicit summation over the vector components $k\in\{x,y,z\}$ is
performed.
Since $\dps\frac{\partial c^{(2)}}{\partial r_k}(\vec{r}-\vec{r'})$ vanishes
for distances $|\vec{r}-\vec{r'}|$ large compared to colloidal length scales
whereas $\vec{w}(\vec{r})$ varies smoothly on colloidal length scales, the
term $w_k(\vec{r})-w_k(\vec{r'})$ in Eq.~\Eq{cdeformed} can be approximated
by
\begin{align}
   w_k(\vec{r})-w_k(\vec{r'}) 
   \approx 
   (r_\ell-r'_\ell)\frac{\partial w_k}{\partial r_\ell}
   \left(\frac{\vec{r}+\vec{r'}}{2}\right) 
   \label{eq:wmw}
\end{align}
with $\ell\in\{x,y,z\}$.
Therefore Eq.~\Eq{cdeformed} can be rewritten as
\begin{multline}
   c^{(2)}(\vec{r}-\vec{r'}+\vec{w}(\vec{r})-\vec{w}(\vec{r'}))
    = c^{(2)}(\vec{r}-\vec{r'})
   \\
   + B_{k\ell}\left(\frac{\vec{r}+\vec{r'}}{2}\right)
   \frac{\partial c^{(2)}}{\partial r_k}(\vec{r}-\vec{r'})(r_\ell-r'_\ell) +
   \mathcal{O}(\|\vec{w}\|^2),
   \label{eq:cdeformed2}
\end{multline}
where the gradient
\begin{align}
   B_{k\ell}(\vec{r}) := \frac{\partial w_k}{\partial r_\ell}(\vec{r})
   \label{eq:defB}
\end{align}
has been introduced.
Using the relation 
\begin{align}
   i(r_\ell-r'_\ell)\exp(i\vec{q}\cdot(\vec{r}-\vec{r'})) =
   \frac{\partial}{\partial q_\ell}\exp(i\vec{q}\cdot(\vec{r}-\vec{r'}))
   \label{eq:relation}
\end{align}
an integration by parts leads to
\begin{align}
   &\frac{\partial c^{(2)}}{\partial r_k}(\vec{r}-\vec{r'})(r_\ell-r'_\ell)
   \notag\\
   &=\int\!\!\frac{\d^3q}{(2\pi)^3}\,iq_k\widehat{\bar{c}}(|\vec{q}|) 
     (r_\ell-r'_\ell)\exp(i\vec{q}\cdot(\vec{r}-\vec{r'}))
   \notag\\
   &=-\!\!\int\!\!\frac{\d^3q}{(2\pi)^3}\,\frac{\partial}{\partial q_\ell}
     \left(q_k\widehat{\bar{c}}(|\vec{q}|)\right)
     \exp(i\vec{q}\cdot(\vec{r}-\vec{r'}))
   \notag\\
   &=-\!\!\int\!\!\frac{\d^3q}{(2\pi)^3}
     \left(\delta_{k\ell}\widehat{\bar{c}}(|\vec{q}|) +
           \frac{q_kq_\ell}{|\vec{q}|}\widehat{\bar{c}}\,'(|\vec{q}|)\right)
     \exp(i\vec{q}\cdot(\vec{r}-\vec{r'})).
   \label{eq:dc2drk}
\end{align}
Noting $B_{kk}(\vec{r})=\nabla\cdot\vec{w}(\vec{r}) = 
\mathcal{O}(\|\vec{w}\|^2)$ (see Eqs.~\Eq{divw} and \Eq{defB}), one obtains
from Eq.~\Eq{cdeformed2}
\begin{multline}
   \bar{c}(|\vec{r}-\vec{r'}+\vec{w}(\vec{r})-\vec{w}(\vec{r'})|)
    = \bar{c}(|\vec{r}-\vec{r'}|)
   \\
   -\!\!\int\!\!\frac{\d^3q}{(2\pi)^3}\,
   \frac{q_kq_\ell}{|\vec{q}|}B_{k\ell}\left(\frac{\vec{r}+\vec{r'}}{2}\right)
   \widehat{\bar{c}}\,'(|\vec{q}|)
   \exp(i\vec{q}\cdot(\vec{r}-\vec{r'}))
   \\
   + 
   \mathcal{O}(\|\vec{w}\|^2).
   \label{eq:cdeformed3}
\end{multline}
Using Eqs.~\Eq{fourierphi} and \Eq{cdeformed3} in Eq.~\Eq{betaFexdeformed}
yields
\begin{align}
   \beta F^\text{ex}[\phi_\vec{w}] 
   &= \beta F^\text{ex}[\phi]
   \notag\\
   &\ \ \ +\frac{1}{2}
    \int\!\!\frac{\d q_z}{2\pi}\!\!\int\!\!\frac{\d p_z}{2\pi}\,
    |q_z|\widehat{B}_{zz}(0,0,p_z)\,\widehat{\bar{c}}\,'(|q_z|)
   \notag\\
   &\ \ \ \hphantom{+ \frac{1}{2}
    \int\!\!\frac{\d q_z}{2\pi}\!\!\int\!\!\frac{\d p_z}{2\pi}}\,
    \widehat{\phi}\left(q_z-\frac{p_z}{2}\right)
    \widehat{\phi}\left(-q_z-\frac{p_z}{2}\right) 
    \notag\\
   &\ \ \ +\mathcal{O}(\|\vec{w}\|^2)
   \label{eq:betaFexdeformed2}
\end{align}
with $\widehat{B}_{zz}(\vec{p})$ being the Fourier transform of 
$B_{zz}(\vec{r})$.
Whereas $\widehat{B}_{zz}(0,0,p_z)$ in Eq.~\Eq{betaFexdeformed2} contributes
only for small wave numbers $p_z$ corresponding to those large length scales
on which the displacement field $\vec{w}(\vec{r})$ varies, 
$\widehat{\bar{c}}\,'(|q_z|)$ contributes only for wave numbers $q_z$ 
corresponding to colloidal length scales.
Consequently $|p_z|\ll|q_z|$ holds for the dominant contributions to the 
integrals in Eq.~\Eq{betaFexdeformed2} so that the approximations
\begin{align}
   \widehat{\phi}\left(\pm q_z-\frac{p_z}{2}\right) \approx
   \widehat{\phi}(\pm q_z)
   \label{eq:approxphi}
\end{align}
apply, thus
\begin{multline}
   \beta F^\text{ex}[\phi_\vec{w}] 
   = \beta F^\text{ex}[\phi]
   \\
   +\frac{1}{2}
   \int\!\!\frac{\d p_z}{2\pi}\,\widehat{B}_{zz}(0,0,p_z)
   \int\!\!\frac{\d q_z}{2\pi}\,|q_z|\,\widehat{\bar{c}}\,'(|q_z|)
   \left|\widehat{\phi}(q_z)\right|^2
   \\
   +\mathcal{O}(\|\vec{w}\|^2).
   \label{eq:betaFexdeformed3}
\end{multline}
Due to Eqs.~\Eq{divw} and \Eq{defB}, the first integral in 
Eq.~\Eq{betaFexdeformed3},
\begin{align}
   &\int\!\!\frac{\d p_z}{2\pi}\,\widehat{B}_{zz}(0,0,p_z)
   \notag\\
   & = \int\d r_x\!\!\int\d r_y\,\frac{\partial w_z}{\partial r_z}(r_x,r_y,0)
   \notag\\
   & = -\int\d r_x\!\!\int\d r_y\,
       \left(\frac{\partial w_x}{\partial r_x}(r_x,r_y,0) +
             \frac{\partial w_y}{\partial r_y}(r_x,r_y,0)\right)
   \notag\\
   &\ \hphantom{=} + \mathcal{O}(\|\vec{w}\|^2)
   \notag\\
   & = -(A_\vec{w} - A) + \mathcal{O}(\|\vec{w}\|^2),
   \label{eq:AwA}
\end{align}
is related to the change $A_\vec{w}-A$ of the cross-sectional area $A$
upon deformation of the container.
This renders the value of the full free energy Eq.~\Eq{betaF} after 
deformation 
\begin{align}
   \beta F[\rho_\vec{w}] 
   =&\!\int\!\d^3r_\vec{w}\,\rho_\vec{w}(\vec{r}_\vec{w})\!
   \left(\ln(\rho_\vec{w}(\vec{r}_\vec{w})\Lambda^3)-1\right)
   \!+\! \beta F^\text{ex}[\phi_\vec{w}]
   \notag\\
   =&\!\int\!\d^3r\,\rho(\vec{r})\left(\ln(\rho(\vec{r})\Lambda^3)-1\right)
   + \beta F^\text{ex}[\phi_\vec{w}]
   \notag\\
   =&\ \beta F[\rho]
    - \frac{A_\vec{w}-A}{4\pi}\int\!\d q_z\,
    |q_z|\,\widehat{\bar{c}}\,'(|q_z|)\left|\widehat{\phi}(q_z)\right|^2
    \notag\\
    &+ \mathcal{O}(\|\vec{w}\|^2).
   \label{eq:betaFdeformed}
\end{align}
Therefore, the interfacial tension $\gamma$ can be expressed as
\begin{align}
   \beta\gamma
   &= \lim\limits_{\|\vec{w}\|\to0}
    \frac{\beta F[\rho_\vec{w}]-\beta F[\rho]}{A_\vec{w}-A}
    \notag\\
   &= -\frac{1}{4\pi}\int\!\d q_z\,
    |q_z|\,\widehat{\bar{c}}\,'(|q_z|)\left|\widehat{\phi}(q_z)\right|^2
    \notag\\
   &= -\frac{1}{2\pi}\int\limits_0^\infty\!\d q_z\,
    q_z\,\widehat{\bar{c}}\,'(q_z)\left|\widehat{\phi}(q_z)\right|^2.
   \label{eq:gamma}
\end{align}
It is worth noting that this expression, according to the derivation given 
above, is independent of the kind of deformations of the cross-sections of the
container perpendicular to the $z$-axis, provided the container is larger
than the size of the colloids such that the displacement field induced by a
deformation of the container walls varies slowly on colloidal length scales.


\subsection{\label{subsec:timedependence}Time-dependence of the interfacial
tension}

Using Eqs.~\Eq{contisolution}--\Eq{timescale} in Eq.~\Eq{gamma} leads to the
time-dependence of the interfacial tension:
\begin{align}
   \beta\gamma(t)
   = -\frac{1}{2\pi}\int\limits_0^\infty\!\d q_z\,
    q_z\,\widehat{\bar{c}}\,'(q_z)\left|\widehat{\phi}(q_z,0)\right|^2
    \exp\left(-\frac{2t}{\tau(q_z)}\right).
   \label{eq:tdgamma}
\end{align}

In the following the example of a step-like initial profile
\begin{align}
   \phi(z,t=0) =
   \begin{cases}
      +\Delta\rho/2 & , z < 0 \\
      -\Delta\rho/2 & , z \geq 0
   \end{cases}
   \label{eq:phi0}
\end{align}
with $\Delta\rho>0$ is considered, which leads to the initial Fourier transform
\begin{align}
   \widehat{\phi}(q_z,t=0) = i\frac{\Delta\rho}{q_z}.
   \label{eq:phihat0}
\end{align}
For this case, Eq.~\Eq{tdgamma} reads
\begin{align}
   \beta\gamma(t)
   = -\frac{\Delta\rho^2}{2\pi}\int\limits_0^\infty\!\d q_z\,
    \frac{\widehat{\bar{c}}\,'(q_z)}{q_z}
    \exp\left(-\frac{2t}{\tau(q_z)}\right),
   \label{eq:tdgammastep}
\end{align}
which allows the determination of the time-dependence of the interfacial
tension $\beta\gamma(t)$ by means of the Fourier transform 
$\widehat{\bar{c}}(q)$ of the direct correlation function.
This procedure is exemplified in the next section for some fluid models.

The particular shape of the initial profile $\phi(z,t=0)$, i.e., whether it is
step-like as in Eq.~\Eq{phi0} or smoothly varying, can be expected to be 
irrelevant for the long-time asymptotic behavior $\beta\gamma(t\to\infty)$.
Using Eq.~\Eq{cq} in Eq.~\Eq{tdgammastep} leads to
\begin{align}
   \beta\gamma(t)
   = -\frac{\Delta\rho^2}{2\pi}\int\limits_0^\infty\!\d q_z\,
    \left(2\widehat{\bar{c}}_2 + \frac{\mathcal{R}'(q_z)}{q_z}\right)
    \exp\left(-\frac{2t}{\tau(q_z)}\right).
   \label{eq:tdgammastepexpansion}
\end{align}
Due to the exponential factor in the integrand, only wave numbers $q_z$ with 
$2t/\tau(q_z)\ll1$ contribute significantly to $\beta\gamma(t)$.
Hence the long-time asymptotics $\beta\gamma(t\to\infty)$ is governed by
wave numbers $q_z$ corresponding to large time scales $\tau(q_z)$.
Since inside the one-phase regions of the phase diagram $0 \leq S(q) \leq 
S(q_\text{max})<\infty$ holds for all $q\in[0,\infty)$, the time scale 
$\tau(q_z)$ defined in Eq.~\Eq{timescale} is large only for small wave numbers
$q_z\to0$.
For $q_z\to0$ the limiting behavior 
\begin{align}
   \left|2\widehat{\bar{c}}_2\right| \gg 
   \left|\frac{\mathcal{R}'(q_z)}{q_z}\right|
   \quad\text{and}\quad
   \tau(q_z)\simeq\frac{S(0)}{Dq_z^2}
   \label{eq:limbehav}
\end{align}
applies, which renders Eq.~\Eq{tdgammastepexpansion}
\begin{align}
   \beta\gamma(t\to\infty)
   &\simeq -\frac{\Delta\rho^2}{2\pi}\int\limits_0^\infty\!\d q_z\,
    2\widehat{\bar{c}}_2\exp\left(-\frac{2Dt}{S(0)}q_z^2\right)
    \notag\\
   &=-\widehat{\bar{c}}_2\Delta\rho^2\sqrt{\frac{S(0)}{8\pi Dt}}.
   \label{eq:tdgammastepasym}
\end{align}
Therefore, at long times $t\to\infty$, the interfacial tension 
$\beta\gamma(t\to\infty)$ is asymptotically proportional to the square 
$\Delta\rho^2$ of the density difference $\Delta\rho$ and inversely 
proportional to the square root $\sqrt{t}$ of the time $t$.
The latter property is of course consistent with the ultimate vanishing of
the interfacial tension due to the disappearance of the interface upon
relaxation.
However, the most interesting feature is that $\beta\gamma(t\to\infty)$ is
proportional to the coefficient $\widehat{\bar{c}}_2$ of the small-$q$
expansion Eq.~\Eq{cq} of the Fourier transform $\widehat{\bar{c}}(q)$.
Depending on the sign of this coefficient $\widehat{\bar{c}}_2$, the interfacial
tension $\beta\gamma(t\to\infty)$ can approach zero with positive or with
negative values, i.e., the relaxation of the colloidal suspension towards a
uniform fluid takes place via an interface of low or of high curvature; high
curvature interfaces typically exhibit fringes (``fingers'').


\section{\label{sec:applications}Applications}

In order to illustrate the general formalism derived in the previous 
Sec.~\ref{sec:formalism}, three examples of model fluids are discussed in
detail here: polymer solutions, charge-stabilized colloids, and colloid-polymer
mixtures.

\subsection{\label{subsec:ps}Polymer solutions}

As the most simple example consider a polymer solution described within the
Gaussian core model \cite{Stillinger1976}, i.e., two polymer chains interact
via the Flory-Krigbaum potential \cite{Flory1950}
\begin{align}
   \beta U(r) = \beta U_0\exp\left(-\frac{r^2}{2d^2}\right)
   \label{eq:gcm}   
\end{align}
with $U_0>0$ measuring the interaction of two polymer chains located at the 
same position and $d$ corresponding to the radius of gyration.
Within the random-phase approximation (RPA) $\bar{c}(r) = -\beta U(r)$ one
obtains 
\begin{align}
   \widehat{\bar{c}}(q_z) 
   &= -\beta U_0(2\pi)^{3/2} d^3 \exp\left(-\frac{(q_zd)^2}{2}\right) \notag\\
   &= -\beta U_0(2\pi)^{3/2} d^3 \left(1 - \frac{(q_zd)^2}{2} + 
      \mathcal{O}\big((q_zd)^4\big)\right),
   \label{eq:cbarps}
\end{align}
hence (see Eq.~\Eq{cq})
\begin{align}
   \widehat{\bar{c}}_0 &= -(2\pi)^{3/2}\beta U_0d^3,
   \label{eq:cbar0ps}\\
   \widehat{\bar{c}}_2 &= \sqrt{2\pi^3}\beta U_0d^5.
   \label{eq:cbar2ps}
\end{align}

\begin{figure}[!t]
   \includegraphics{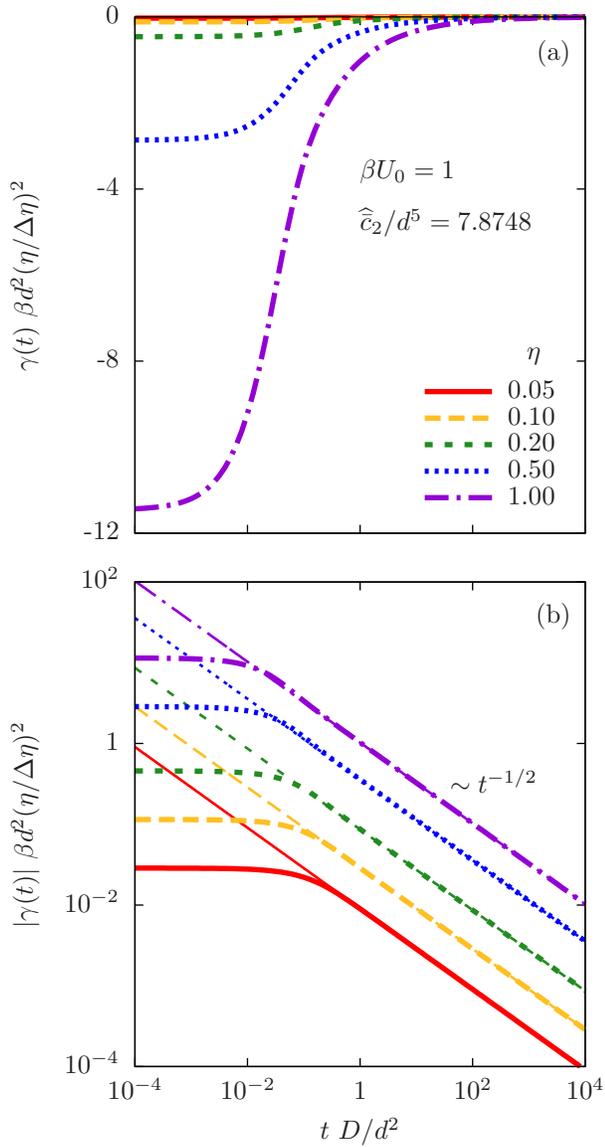}
   \caption{Non-equilibrium interfacial tension $\gamma(t)$ 
            during the relaxation of polymer solutions described by a Gaussian
            core model Eq.~\Eq{gcm} with polymer number densities $\bar{\rho}$
            corresponding to packing fractions 
            $\eta=\pi\bar{\rho}d^3/6\in\{0.05,0.1,0.2,0.5,1\}$.
            Thick lines represent the full expression Eq.~\Eq{tdgammastep},
            whereas thin lines in panel (b) correspond to the long-time
            asymptotics Eq.~\Eq{tdgammastepasym}.
            Panel (a) shows that, for this system, the interfacial tension is 
            negative, $\gamma(t)<0$, for all times $t\geq0$.
            From panel (b) one can infer that the asymptotic expression 
            Eq.~\Eq{tdgammastepasym} applies for times $t \gtrsim d^2/D$.}
   \label{fig:2}
\end{figure}

Figure~\ref{fig:2} displays the temporal evolution of the interfacial tension
$\gamma(t)$ for the case of interaction strength $\beta U_0=1$ and packing
fractions $\eta:=\pi\bar{\rho}d^3/6\in\{0.05,0.1,0.2,0.5,1\}$.
The thick lines correspond to the full expression Eq.~\Eq{tdgammastep} based
on the direct correlation function Eq.~\Eq{cbarps}, whereas the thin lines
in Fig.~\ref{fig:2}(b) are the long-time asymptotics Eq.~\Eq{tdgammastepasym}.
Note that, according to Eq.~\Eq{tdgammastep}, the interfacial tension scales
$\sim(\Delta\eta/\eta)^2$ with the initial relative density difference 
$\Delta\rho/\rho=\Delta\eta/\eta$.

The interfacial tension $\gamma(t)$ for this system turns out to be negative
for all times $t\geq0$ (see Fig.~\ref{fig:2}(a)) because 
$\widehat{\bar{c}}(q_z)$ in Eq.~\Eq{cbarps} is a monotonically increasing 
function of $q_z$, i.e., $\widehat{\bar{c}}\,'(q_z)>0$, so that, according to
Eq.~\Eq{tdgammastep}, $\gamma(t)<0$.
This is, in particular, in accordance with Eq.~\Eq{tdgammastepasym} and
$\widehat{\bar{c}}_2>0$ from Eq.~\Eq{cbar2ps}. 
Figure~\ref{fig:2}(b) indicates that the asymptotic behavior $\gamma(t)\sim 
t^{-1/2}$ in Eq.~\Eq{tdgammastepasym} can be expected to apply at times beyond
the Brownian time scale $t \gtrsim d^2/D$; this statement applies to a wide 
range of the parameters $\beta U_0$ and $\eta$, the results of which are not
shown here.

According to Refs.~\cite{Kim1990, Francis2006}, the interaction potential
Eq.~\Eq{gcm} with interaction strength $\beta U_0 = 1.58$ and
$d = R_g\sqrt{2/3}=60\,\text{nm}$ is valid to describe the interaction of
pairs of poly($\alpha$-methyl styrene) chains with molecular mass 
$M_w=2.96\times10^6\,\text{g/mol}$ in toluene at $25\,^\circ\text{C}$ in the
dilute regime $\eta \lesssim 0.22$.
At the Brownian time $t=d^2/D=46\,\text{ms}$, where 
$D=0.079\,\mu\text{m}^2/\text{s}$ is the experimental diffusion constant, one
finds $\beta\gamma(t)d^2=-0.13(\Delta\eta/\eta)^2$ for $\eta=0.22$, i.e., for
an initial perturbation of $|\Delta\eta|=0.1\eta$ the free energy 
$k_\text{B}T$ is gained by interface fluctuations which increase the 
interfacial area by $2.77\,\mu\text{m}^2$. 


\subsection{\label{subsec:csc}Charge-stabilized colloids}

As a second example consider a colloidal suspension of charge-stabilized
colloids of diameter $d$ in a solvent with inverse Debye length $\kappa$.
Here the interaction of two colloidal particles is modeled by the potential
\begin{align}
   \beta U(r) =
   \begin{cases}
      \infty                                       & , r \leq d \\
      \dps\beta U_d \frac{\exp(-\kappa(r-d))}{r/d} & , r > d,
   \end{cases}
   \label{eq:Ucsc}
\end{align}
where the strength $\beta U_d>0$ of the double layer repulsion increases with
the charge of the colloids.
The direct correlation function of the colloidal suspension within RPA of
the double layer repulsion in excess to the hard-sphere reference system is
given by
\begin{align}
   \bar{c}(r) =
   \begin{cases}
      c_\text{HS}(r)                                & , r \leq d \\
      \dps-\beta U_d \frac{\exp(-\kappa(r-d))}{r/d} & , r > d,
   \end{cases}
   \label{eq:ccsc}
\end{align}
where the Percus-Yevick direct correlation function of hard spheres with 
diameter $d$ and packing fraction $\eta$ is \cite{Hansen1986}
\begin{align}
   c_\text{HS}(r) =
   \begin{cases}
      \dps c_\text{HS}^{(0)}(\eta)+c_\text{HS}^{(1)}(\eta)\frac{r}{d}+
      c_\text{HS}^{(3)}(\eta)\left(\frac{r}{d}\right)^3 & , r \leq d \\
      0                                                 & , r > d
   \end{cases}
   \label{eq:py}
\end{align} 
with
\begin{align}
   c_\text{HS}^{(0)}(\eta) &= -\frac{(1+2\eta)^2}{(1-\eta)^4}, \notag\\
   c_\text{HS}^{(1)}(\eta) &= \frac{6\eta(1+\eta/2)^2}{(1-\eta)^4}, \notag\\
   c_\text{HS}^{(3)}(\eta) &= -\frac{\eta(1+2\eta)^2}{2(1-\eta)^4}. 
   \label{eq:coeffpy}
\end{align}

\begin{figure}[!t]
   \includegraphics{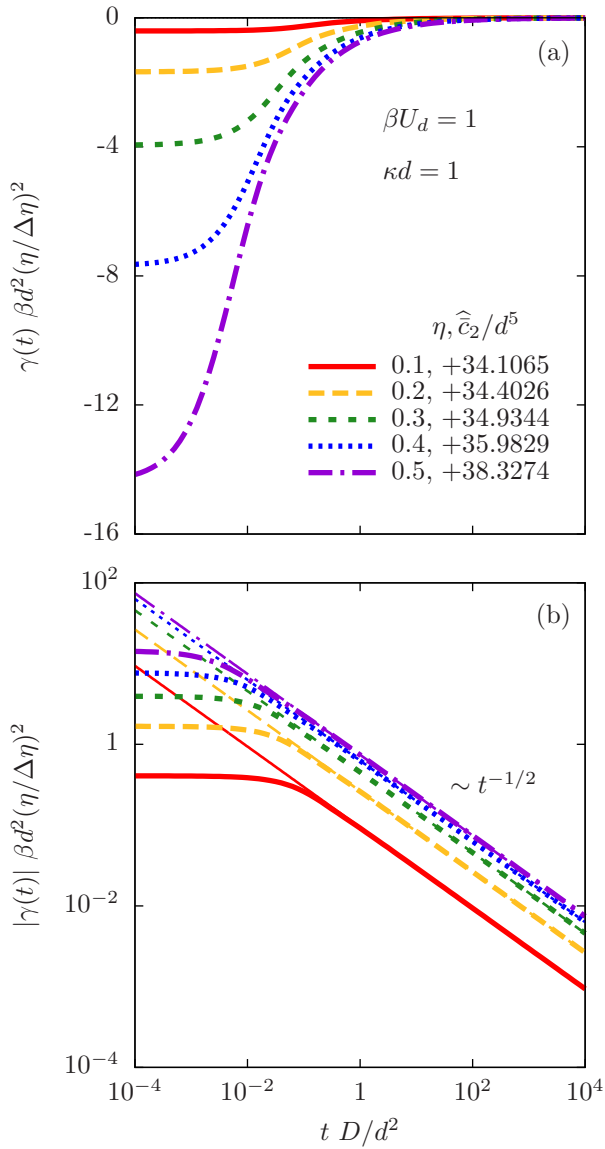}
   \caption{Non-equilibrium interfacial tension $\gamma(t)$ 
            during the relaxation of suspensions of charge-stabilized colloids
            interacting via the potential Eq.~\Eq{Ucsc} with strength $\beta 
            U_d=1$ of the double layer repulsion, inverse Debye length $\kappa
            d=1$, and packing fractions $\eta\in\{0.1,0.2,0.3,0.4,0.5\}$.
            Thick lines represent the full expression Eq.~\Eq{tdgammastep},
            whereas thin lines in panel (b) correspond to the long-time
            asymptotics Eq.~\Eq{tdgammastepasym}.
            Panel (a) shows that, for this system, the interfacial tension is 
            negative, $\gamma(t)<0$, for all times $t\geq0$.
            From panel (b) one can infer that the asymptotic expression 
            Eq.~\Eq{tdgammastepasym} applies for times $t \gtrsim d^2/D$.}
   \label{fig:3}
\end{figure}

Figure~\ref{fig:3} displays the temporal evolution of the interfacial tension
$\gamma(t)$ for the case of interaction strength $\beta U_d=1$, inverse Debye
length $\kappa d=1$, and packing fractions $\eta\in\{0.1,0.2,0.3,0.4,0.5\}$.
The thick lines correspond to the full expression Eq.~\Eq{tdgammastep} based
on the direct correlation function Eqs.~\Eq{ccsc}--\Eq{coeffpy}, whereas the 
thin lines in Fig.~\ref{fig:3}(b) are the long-time asymptotics 
Eq.~\Eq{tdgammastepasym}.

The interfacial tension $\gamma(t)$ for charge-stabilized colloids, as for
polymer solutions in the previous Subsec.~\ref{subsec:ps}, is negative, 
$\gamma(t)<0$, for all times $t\geq0$ (see Fig.~\ref{fig:3}(a)), which is
also in agreement with the positive values $\widehat{\bar{c}}_2>0$ entering
in the long-time asymptotics Eq.~\Eq{tdgammastepasym}.
Figure~\ref{fig:3}(b) indicates that the asymptotic behavior $\gamma(t)\sim 
t^{-1/2}$ in Eq.~\Eq{tdgammastepasym} sets in, as for polymer solutions in 
the previous Subsec.~\ref{subsec:ps}, at times beyond the Brownian time 
scale $t \gtrsim d^2/D$.

It has to be noted that the interaction strength $\beta U_d$ of some real 
charge-stabilized colloids can exceed unity by orders of magnitudes: In 
Ref.~\cite{Crocker1994}, e.g., an aqueous dispersion 
($\kappa^{-1}=161\,\text{nm}$) of polystyrene sulfate spheres of diameter 
$d=652\,\text{nm}$ and packing fraction $\phi\approx0.04$ is reported with 
$\beta U_d=463$.
For such a large value of $\beta U_d$ the RPA Eq.~\Eq{ccsc} is not justified
so that one has to use more reliable schemes to evaluate $\bar{c}(r)$.


\subsection{\label{subsec:cpm}Colloid-polymer mixtures}

As a final example consider a colloid-polymer mixture composed of colloidal 
hard spheres of diameter $d$ suspended in a solution of polymer coils of 
diameter $\ell$.
Within the Asakura-Oosawa-Vrij theory \cite{Asakura1954,Asakura1958,Vrij1976},
the polymer coils with packing fraction $\eta_\text{p}$ give rise to an 
effective depletion interaction 
\begin{align}
   \beta V(r) = -\eta_\text{p}\left(1+\frac{d}{\ell}\right)^3
   \left(1 - \frac{3r}{2(d+\ell)} + \frac{r^3}{2(d+\ell)^3}\right)
   \label{eq:depletion}
\end{align}
between colloidal particles for distances in the range $r\in(d,d+\ell)$.
The direct correlation function of the colloidal suspension within RPA of
the depletion interaction in excess to the hard-sphere reference system is 
given by
\begin{align}
   \bar{c}(r) =
   \begin{cases}
      c_\text{HS}(r) & , r \leq d \\
      -\beta V(r)    & , r \in (d,d+\ell) \\
      0              & , r \geq d+\ell.
   \end{cases}
   \label{eq:ccolloids}
\end{align}

\begin{figure}[!t]
   \includegraphics{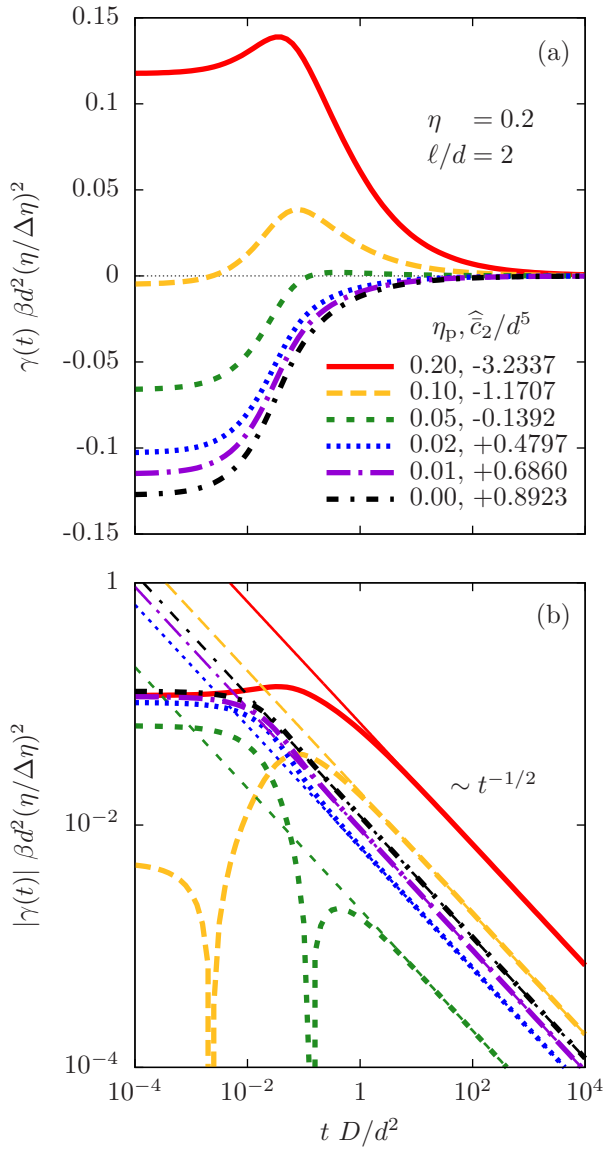}
   \caption{Non-equilibrium interfacial tension $\gamma(t)$ 
            during the relaxation of colloid-polymer mixtures described within
            the Asakura-Oosawa-Vrij model Eq.~\Eq{depletion} with colloidal 
            packing fraction $\eta=0.2$, polymer-to-colloid size ratio 
            $\ell/d=2$ and polymer packing fractions $\eta_p\in\{0,0.01,0.02,
            0.05,0.1,0.2\}$.
            Thick lines represent the full expression Eq.~\Eq{tdgammastep},
            whereas the thin lines in panel (b) correspond to the long-time
            asymptotics Eq.~\Eq{tdgammastepasym}.
            Panel (a) demonstrates that the interfacial tension $\gamma(t)$ as
            a function of time $t$ can be positive ($\gamma(t)>0$, see case 
            $\eta_p=0.2$) or negative ($\gamma(t)<0$, see cases $\eta_p\in
            \{0,0.01,0.02\}$) or it can change its sign (see cases $\eta_p
            \in\{0.05,0.1\}$). 
            From panel (b) one can infer that the asymptotic expression 
            Eq.~\Eq{tdgammastepasym} applies for times $t \gtrsim d^2/D$.}
   \label{fig:4}
\end{figure}

Figure~\ref{fig:4} displays the temporal evolution of the interfacial tension
$\gamma(t)$ for the colloidal packing fraction $\eta=0.2$, the size ratio
$\ell/d=2$, and the polymer packing fractions $\eta_\text{p}\in\{0,0.01,0.02,
0.05,0.1,0.2\}$.
The thick lines correspond to the full expression Eq.~\Eq{tdgammastep} based
on the direct correlation function Eq.~\Eq{ccolloids}, whereas the thin
lines in Fig.~\ref{fig:4}(b) are the long-time asymptotics 
Eq.~\Eq{tdgammastepasym}.

In accordance with Eq.~\Eq{tdgammastepasym}, the asymptotic decay 
$\gamma(t)\to0$ is from below for weak depletion attractions, $\eta_\text{p}
\in\{0,0.01,0.02\}$, with positive values $\widehat{\bar{c}}_2>0$, whereas it
is from above for strong depletion attractions, $\eta_\text{p}\in\{0.05,0.1,
0.2\}$, with negative values $\widehat{\bar{c}}_2<0$.
This demonstrates that it can be possible to alter the sign of the asymptotic
decay by simply changing the polymer concentration. 
Moreover, the cases $\eta_p\in\{0.05,0.1\}$ in Fig.~\ref{fig:4} show that the
sign of the interfacial tension $\gamma(t)<0$ at early times $t\to0$ has
no influence on the sign $\gamma(t)>0$ at late times $t\to\infty$.
Figure~\ref{fig:4}(b) indicates that also in these systems the asymptotic 
behavior $\gamma(t)\sim t^{-1/2}$ in Eq.~\Eq{tdgammastepasym} can be expected
to apply at times beyond the Brownian time scale $t \gtrsim d^2/D$.


\section{\label{sec:conclusions}Conclusions}

It has been shown in Sec.~\ref{sec:formalism} that, under the mild conditions
of an interaction potential $U(r)$ decaying for $r\to\infty$ faster than $\sim
r^s$ with $s>5$ (see Subsec.~\ref{subsec:colloids}) and of an underlying model
B dynamics (see Subsec.~\ref{subsec:ddft}), an initially perturbed 
colloidal
fluid relaxes towards equilibrium accompanied by a time-dependent 
non-equilibrium interfacial tension $\gamma(t)$ which is not necessarily 
positive.

Here
the non-equilibrium interfacial tension $\gamma(t)$ is defined via the work
required to deform the system such that the interfacial area changes while 
the fluid volume is preserved (see Subsec.~\ref{subsec:tension}).
An experimental determination of the non-equilibrium interfacial tension 
$\gamma(t)$ requires, on the one hand, a sufficiently slow dynamics of the
system such that the structure does not change appreciably during the 
measurement and, on the other hand, a sufficiently fast dynamics to preserve
local equilibrium during deformation.
Colloidal suspensions and polymer solutions with their wide separation of 
solvent and solute time scales (nanoseconds vs.\ milliseconds) are systems 
for which the negative non-equilibrium interfacial tension could be realized.
There the fluid, whose non-equilibrium interfacial tension is to be measured,
is formed by the colloidal particles, whereas the solvent acts as a medium
whose Stokes flow due to agitation of the container walls leads to the 
deformation of the colloidal fluid. 

The non-equilibrium interfacial tension $\gamma(t)$ is, under the 
above-mentioned conditions, determined by the equilibrium structure, the
diffusion constant, and the initial density difference $\Delta\rho$ from
equilibrium (see Subsec.~\ref{subsec:timedependence}).
In particular, Eq.~\Eq{tdgammastep} leads to $\gamma(t)\sim\Delta\rho^2$, 
which is reminiscent of the expression for the interfacial tension of 
equilibrium interfaces within the square-gradient approximation 
\cite{Rowlinson2002}.
However, the square-gradient approximation is applicable only, if the 
square-gradient contribution, and hence the interfacial tension, is positive,
because otherwise all uniform bulk states would be unstable with respect to 
density fluctuations.
A positive square-gradient contribution corresponds to a negative coefficient
$\widehat{\bar{c}}_2<0$ of the quadratic term $\sim q^2$ in Eq.~\Eq{cq}, but
this term is not required to be negative beyond square-gradient approaches,
where the higher-order terms $\mathcal{R}(q)$ are not neglected.
The interesting and rather general finding Eq.~\Eq{tdgammastepasym} shows 
that the long-time limit $\gamma(t\to\infty)$ is proportional to this 
coefficient $\widehat{\bar{c}}_2$ of the quadratic contribution $\sim q^2$ in 
Eq.~\Eq{cq}.
The reason is that the relaxation time $\tau(q)$ in Eq.~\Eq{timescale} 
decreases with increasing wave number $q$ so that the contributions of large
wave numbers $q$ decay quickly (see Eq.~\Eq{tdgammastep}).

Several examples of realistic systems have been proposed in 
Sec.~\ref{sec:applications} for which negative non-equilibrium interfacial 
tensions $\gamma(t)$ during relaxation can be expected to occur: Polymer
solutions (Subsec.~\ref{subsec:ps}, Fig.~\ref{fig:2}), charge-stabilized
colloids (Subsec.~\ref{subsec:csc}, Fig.~\ref{fig:3}), and colloid-polymer
mixtures (Subsec.~\ref{subsec:cpm}, Fig.~\ref{fig:4}).
The latter system even offers the possibility to switch between an
asymptotically positive ($\gamma(t\to\infty)>0$) and an asymptotically 
negative ($\gamma(t\to\infty)<0$) non-equilibrium interfacial tension.
As a rule of thumb, $\widehat{\bar{c}}_2<0$, and hence $\gamma(t\to\infty)>0$,
corresponds to an interaction potential $U(r)$ with the attractive 
contribution being sufficiently strong as compared to the repulsive 
contribution, whereas $\widehat{\bar{c}}_2>0$, and hence $\gamma(t\to\infty)
<0$, is the result of an interaction potential $U(r)$ with the repulsive
contribution being sufficiently strong as compared to the attractive
contribution.

Positive and negative values of the non-equilibrium interfacial tension 
$\gamma(t)$ lead to different morphologies of density perturbations:
The relaxation towards a uniform equilibrium state takes place by minimizing
the interfacial area for $\gamma(t)>0$ and by maximizing the interfacial area
for $\gamma(t)<0$.
Hence, localized density perturbations tend to spatially shrink with interfaces
being smooth for $\gamma(t)>0$ while they tend to spread out with increasingly
rough interfaces for $\gamma(t)<0$.

It is important to realize that the relaxation of a density perturbation is
not driven by the non-equilibrium interfacial tension,
which can be positive or negative (Fig.~\ref{fig:4}(a)),
but by the non-uniformity of the local chemical potential (see 
Eq.~\Eq{mudef}); the 
interfacial tension is merely related to the interfacial structure formed 
during the relaxation process.
It is straightforward to show with Eqs.~\Eq{mudef}, \Eq{jdef}, and 
\Eq{contirho} that the Helmholtz free energy $F[\rho(t)]$ is a
monotonically decreasing function of time $t$,
\begin{align}
   \frac{\d F[\rho(t)]}{\d t} = 
   -\frac{1}{\beta D}\int\d^3r\,\frac{\vec{j}(\vec{r},t)^2}{\rho(\vec{r},t)}
   < 0,
\end{align}
and it is this decrease which leads to the irreversibility of the relaxation
towards equilibrium.
In contrast, the non-equilibrium interfacial tension in Eq.~\Eq{gamma} 
quantifies the linear response of the system to reversible deformations, and
it has been shown in Sec.~\ref{subsec:tension} that this linear response
is independent of the type of deformation.
Beyond linear response one may find a dependence of the system's response on
the type of deformation, but such a behavior has to be described by a different
quantity than the interfacial tension.

Based on these comments, one can reconcile the non-equilibrium interfaces during
relaxation studied in the present work with equilibrium interfaces between two
coexisting bulk phases: First, equilibrium interfaces, whose interfacial 
tensions do not vanish, exhibit a time-independent structure because the local
chemical potential in equilibrium is uniform.
Second, bulk coexistence is possible only for a sufficiently strong attractive
contribution to the interaction potential $U(r)$ as compared to the repulsive
contribution, which typically leads to $\widehat{\bar{c}}_2<0$, i.e., 
$\gamma(t\to\infty)>0$. 
Conversely, a negative equilibrium interfacial tension requires an interaction
potential $U(r)$ with the repulsive contribution dominating the attractive one,
but under these condition no bulk coexistence, and thus no equilibrium 
interface, occurs.

In conclusion, a rather general expression for the interfacial tension of a
relaxing planar non-equilibrium interface is derived and applied to realistic
systems of colloidal suspensions and polymer solutions.
It is shown that the non-equilibrium interfacial tension is not necessarily
positive, that negative non-equilibrium interfacial tensions are consistent
with strictly positive equilibrium interfacial tensions, and that the sign of
the interfacial tension can influence the morphology of density perturbations
during relaxation.
The present study highlights that the useful concept of a non-equilibrium 
interfacial tension shares some but not all properties with the equilibrium
interfacial tension.
Until now concepts of non-equilibrium interfacial tensions have been introduced
only for systems close to equilibrium, whose known relaxation dynamics 
towards equilibrium allows for generalizations of the equilibrium interfacial
tension.
It is a future task of enormous relevance to understand the properties of
non-equilibrium interfaces also far away from equilibrium.
Whether the interfacial tension is a useful concept also far away from 
equilibrium or whether its applicability is restricted to the vicinity of
equilibrium is an interesting open question.





\end{document}